\documentstyle[12pt]{article}
\begin{document}
\baselineskip=23truept
\begin{center}
{\large\bf Title: Critical curvature of large-$N$ nonlinear
$O(N)$ sigma model on $S^2$}\\
{\bf Short title: Critical curvature on $S^2$}
\end{center}
\vspace{1cm}
\begin{center}
K.H. Kim and Dae-Yup Song\footnote{Electronic address:
	  dsong@sunchon.sunchon.ac.kr}
\end{center}
\begin{center}
{\it Department of Physics, Sunchon National University,\\
Sunchon, Chonnam 540-742, Korea }
\end{center}
\bigskip
\begin{center}Abstract\end{center}
We study the nonlinear $O(N)$ sigma model on $S^2$ with the
gravitational coupling term, by evaluating the effective potential in
the large-$N$ limit. It is shown that there is
a critical curvature $R_c$ of $S^2$ for any positive gravitational
coupling constant $\xi$, and the dynamical mass generation
takes place only when $R<R_c$. The critical curvature is analytically
found as a function of $\xi$ $(>0)$, which leads us to define a function
looking like a natural generalization of
Euler-Mascheroni constant.


\newpage

\section{Introduction}

For a model described by boson fields $n_i$ $(n=1,2,\cdots,N)$
on a curved spacetime, a gravitational interaction term $\xi_i R n_i n_i$
may be added to the Lagrangian \cite{BiDa}, where $\xi_i$ is a gravitational
coupling constant and $R$ is the scalar curvature of the spacetime
( In this paper, we consider only Euclidean spacetime. ). Since
this interaction term is a bilinear term of the fields, the change of
mass through quantum loop corrections may be of interest.

Recently one of us \cite{Song} studied the three dimensional nonlinear
$O(N)$ sigma model on $S^2\times R^1$ in the leading order
of $1/N$ expansion,
where $S^2$ is the two sphere with the canonical metric. For the $O(N)$
symmetry we assume that $\xi_i$ is the same $\xi$ for all $i$. On
the three dimensional flat spacetime, there exists a critical
coupling constant of the model and the dynamical mass generation
takes place only when the
coupling constant is larger than the critical one (strong coupling
regime) \cite{Aref}. In the strong coupling regime, we showed that
there is a
critical curvature (radius) of $S^2$ for some value of $\xi$. Though
the $O(N)$ model can be treated by the $1/N$ expansion in both two and
three dimensions, the properties of the models are substantially
different. One of them is that the dynamical mass generation takes
place in two dimension no matter how small the coupling constant is,
as it should be.

In this paper, we will study the two-dimensional $O(N)$ model
\cite{PolBerZin} on $S^2$ with the gravitational coupling term
in the large-$N$ limit, by evaluating the effective potential
\cite{Colman}. We will show that there is a critical curvature $R_c(\xi)$ for
every $\xi~(>0)$. As $\xi$ approaches 0, the $R_c$ goes to infinity
and decreases monotonically as $\xi$ increases. By introducing a function
which looks like a functional generalization of Euler-Mascheroni constant,
the $R_c$ can be neatly expressed by the function.

\section{Effective potential and critical curvature}
We assume that the
metric of $S^2$ is the canonical one so that the curvature $R$ is
$2/\rho^2$, where $\rho$ is the radius of $S^2$. Then the
Lagrangian of our model is written as:
\begin{eqnarray}
L=\int d\theta d\varphi ~\rho^2 \sin\theta
  +(\rho^2\sin^2\theta)^{-1}\partial_\varphi n^i \partial_\varphi n^i
  \nonumber\\
 &&~+\xi R n^2 +\sigma(n^2-Ng^{-2})~],
\end{eqnarray}
where $g_0$ is the bare coupling constant.
For the evaluation of the effective potential we can write the
Lagrangian density as
\begin{equation}
{\cal L}=n^i Dn_i-N\sigma/g_0^2.
\end{equation}
In (2)  $D$ is defined as $\rho^{-2}({\bf L}^2+2\xi)+\sigma$, where
${\bf L}$ is the quantum mechanical angular momentum operator.
By performing the Gaussian functional integral \cite{Jackiw}, the effective
potential per unit volume is given as
\begin{equation}
\frac{V}{N}=-\frac{\sigma}{g_0^2}+\frac{1}{4\pi\rho^2}({\rm Tr} \ln D+ C),
\end{equation}
where $C$ is a constant which may be determined by requiring
$V\mid_{\sigma=0}=0$. As is well-known, in the leading order of $1/N$
expansion the effective potential in (3) is written in terms of spacetime
independent $\sigma$:
\begin{equation}
\frac{V}{N}=-\frac{\sigma}{g_0}+\frac{1}{4\pi\rho^2}\sum_{l=0}^{I}
   (2l+1)\ln(1+\frac{\sigma}{\frac{l(l+1)}{\rho^2}+\frac{2\xi}{\rho^2} }),
\end{equation}
where we introduce an integer cut-off $I$ for the quantum
number of the operator ${\bf L}$.

In order to compare effective potential in (4)  with that on
$R^2$, as in the previous paper \cite{Song}, one can use the following
formula:
\[
\sum_{l=0}^N f(l)=\frac{1}{2}f(0)+\int_0^{N+1}f(x)dx
		  +\sum_{l=0}^N\int_0^1 f'(x+l)(x-\frac{1}{2})dx
		  -\frac{1}{2}f(N+1),
\]
which gives
\begin{eqnarray}
\frac{V}{N}
&=&-\frac{\sigma}{g_0^2}
  +\frac{1}{2\pi \rho} \int_0^I(x+\frac{1}{2})
  \ln (1+\frac{\sigma \rho^2}{(x+\frac{1}{2})^2+2\xi-\frac{1}{4}})dx
  +\frac{1}{8\pi\rho^2} \ln (1+\frac{\sigma \rho^2}{2\xi}) \nonumber \\
&&+\frac{1}{2\pi \rho^2}\sum_{l=0}^I\int_0^1 dx (x-\frac{1}{2})
  \left[ \begin{array}{l}
	\ln (1+\frac{\sigma\rho^2}{(x+l+\frac{1}{2})^2+2\xi-\frac{1}{4}})\\
	+ \frac{2(x+l+\frac{1}{2})^2}{(x+l+\frac{1}{2})^2
	+2\xi- \frac{1}{4} +\sigma \rho^2}
	- \frac{2(x+l+\frac{1}{2})^2}{(x+l+\frac{1}{2})^2 +2\xi- \frac{1}{4}}
   \end{array}\right].
\end{eqnarray}
Since the integrals in (5) are well defined without infra-red
problem for $\xi>0$, we will restrict our attention to these cases.
By defining a cut-off $\Lambda=(I+\frac{1}{2})/\rho$ whose dimension
is that of a momentum, we can write the potential as
\begin{eqnarray}
\frac{V}{N}&=& -\frac{\sigma}{g_0^2}
	 +\frac{1}{2\pi}\int_{1/2\rho}^\Lambda y
	 \ln (1+\frac{\sigma}{y^2+\frac{2\xi-\frac{1}{4}}{\rho^2}})dy
	 +\frac{1}{8\pi \rho^2}\ln (1+\frac{\sigma \rho^2}{2\xi}) \nonumber\\
	&&+\frac{1}{2\pi \rho^2}\sum_{l=1}^\infty
	   \int_{-\frac{1}{2}}^{\frac{1}{2}} dt~t
	   \left[\begin{array}{l}
	   \ln (1+\frac{\sigma\rho^2}{(t+l)^2+2\xi -\frac{1}{4}})\\
	   +\frac{2(t+l)^2}{(t+l)^2+2\xi-\frac{1}{4}+\sigma\rho^2}
	   -\frac{2(t+l)^2}{(t+l)^2+2\xi-\frac{1}{4}}
	   \end{array}\right] \nonumber\\
	&&+O(1/\Lambda) \nonumber \\
 &=&-\frac{\sigma}{g_0^2}-\frac{\sigma}{4\pi}
	 (\ln \frac{2\xi+\sigma\rho^2}{\rho^2\Lambda^2} -1)
	 +\frac{1-4\xi}{8\pi\rho^2}\ln (1+\frac{\sigma\rho^2}{2\xi})\nonumber\\
	&&+\frac{1}{2\pi \rho^2}\sum_{l=1}^\infty
	   \int_{-\frac{1}{2}}^{\frac{1}{2}} dt~t
	   \left[\begin{array}{l}
	   \ln (1+\frac{\sigma\rho^2}{(t+l)^2+2\xi -\frac{1}{4}})\\
	   +\frac{2(t+l)^2}{(t+l)^2+2\xi-\frac{1}{4}+\sigma\rho^2}
	   -\frac{2(t+l)^2}{(t+l)^2+2\xi-\frac{1}{4}}
	   \end{array}\right] \nonumber\\
	&&+O(1/\Lambda).
\end{eqnarray}
The effective potential in (6) once again confirms the well-known
fact that topological change of spacetime does not give rise to
new counterterms, and thus we can use the renormalization relation
of the model on $R^2$ \cite{Colman}
\begin{equation}
-\frac{1}{g_0^2} =-\frac{1}{g^2} + \frac{1}{4\pi}\ln \frac{M^2}{\Lambda^2},
\end{equation}
where $M$ is the renormalization mass.
It is convenient to define  $\sigma_0$
\begin{equation}
\sigma_0=M^2e^{-4\pi/g^2}
\end{equation}
which denotes the square of the dynamically generated mass on $R^2$.
Making use of the relations (7) and (8), one can find the renormalized
effective potential
\begin{eqnarray}
\frac{V}{N}&=& -\frac{\sigma}{4\pi}
	(\ln\frac{2\xi +\sigma\rho^2}{ \sigma_0 \rho^2}-1)
	+\frac{1-4\xi}{8\pi\rho^2}\ln (1+\frac{\sigma\rho^2}{2\xi})
	\nonumber\\
  &&+\frac{1}{2\pi \rho^2}\sum_{l=1}^\infty
	\int_{-\frac{1}{2}}^{\frac{1}{2}} dt~t
	 \left[\begin{array}{l}
	   \ln (1+\frac{\sigma\rho^2}{(t+l)^2+2\xi -\frac{1}{4}})\\
	   +\frac{2(t+l)^2}{(t+l)^2+2\xi-\frac{1}{4}+\sigma\rho^2}
	   -\frac{2(t+l)^2}{(t+l)^2+2\xi-\frac{1}{4}}
	   \end{array}\right].
\end{eqnarray}
To have a better understanding of the shape of $V$ in (9),
we evaluate the first derivative of $V$ with respect to $\sigma$:
\begin{eqnarray}
\frac{1}{N}\frac{\partial V}{\partial \sigma}\equiv\frac{V'(\sigma)}{N}
&=&-\frac{1}{4\pi}\ln\frac{2\xi+\sigma\rho^2}{\sigma_0\rho^2}
   +\frac{1}{8\pi}\frac{1}{2\xi+\sigma\rho^2}    \nonumber\\
& &+\frac{1}{2\pi}\sum_{l=1}^\infty \int_{-\frac{1}{2}}^\frac{1}{2}
   dt~t\frac{-(t+l)^2+2\xi-\frac{1}{4}+\sigma\rho^2}
		{[(t+l)^2+2\xi-\frac{1}{4}+\sigma\rho^2]^2}.
\end{eqnarray}
As shown in the appendix in detail, one can find a simpler form of the first
derivative in (10):
\begin{equation}
\frac{1}{N}\frac{\partial V}{\partial \sigma}
=\frac{1}{4\pi}[\ln(\sigma_0\rho^2)+\frac{1}{2\xi+\sigma\rho^2}
+2\gamma(2\xi+\sigma\rho^2-\frac{1}{4})],
\end{equation}
where the function $\gamma$ is defined by
\begin{equation}
\gamma(\beta)=\lim_{N\rightarrow\infty}
   [\sum_{n=1}^N\frac{n+\frac{1}{2}}{(n+\frac{1}{2})^2+\beta}
	-\frac{1}{2}\ln(N^2+\beta)].
\end{equation}
The $\gamma$ function looks like a generalization of Euler-Mascheroni
constant
\[\gamma_E=\lim_{N\rightarrow\infty}[\sum_{n=1}^N n^{-1} -\ln N]
   =0.5772\cdots.\]
In fact, making use of the formulae e.g. in \cite{WhiWat}, one can show that
$\gamma(0)=\gamma_E-2-2\ln2$. Furthermore, it is easy to show that
$\gamma(\beta)$ $(\beta>-\frac{1}{4})$  is a monotonically decreasing
function of $\beta$, and it approaches to $-\infty$ as $\beta$ goes to
$\infty$.

Now one can find that, for a fixed $\xi$, $V'(\sigma)$ monotonically
decreases to $-\infty$ as $\sigma$ increases. That is, a global
stationary point of the effective potential $V$ which denotes
the dynamical mass generation appears only when $V'(0) >0$.
Therefore, the critical radius of $S^2$ is given by
\begin{equation}
\sigma_0\rho_c^2=\exp(-\frac{1}{2\xi}-2\gamma(2\xi-\frac{1}{4})).
\end{equation}
Or, the critical curvature is given by
\begin{equation}
R_c=2\sigma_0\exp(\frac{1}{2\xi}+2\gamma(2\xi-\frac{1}{4})).
\end{equation}
The critical curvature which is $\infty$ in the limit
$\xi\rightarrow 0$ decreases monotonically to 0 as $\xi$ increases
to $\infty$, and the dynamical mass generation takes place only when
$R<R_c$ (or $\rho>\rho_c$).

\section{Discussion}

By evaluating the  effective potential (9) in the leading order of $1/N$
expansion, we have shown that,
the $O(N)$ nonlinear $\sigma$ model with the gravitational coupling
term described by the Lagrangian (1), has a critical curvature ( given in
(13,14) ) which decreases as $\xi$ increases.

For the conformal symmetry, $\xi$ must be $\xi_c=\frac{1}{4}[(d-2)/(d-1)]$
in a $d$-dimensional spacetime. That is, the conformal coupling constant
$\xi_c$ is 0  in two dimension and $\frac{1}{8}$ in three dimension.
One of the common features of the $O(N)$ nonlinear $\sigma$ model
on $S^2$ and $S^2\times R^1$ \cite{Song} is that the dynamical mass
generation takes place for any finite $R$ in the limit
$\xi\rightarrow\xi_c$ while there exists a critical curvature for
$\xi>\xi_c$.
\bigskip
\begin{center}
{\bf Acknowledgments}
\end{center}
One of us (D.Y.S) would like to thank Jeong Hyeong Park for the discussion
on the $\gamma$ function. This work is supported in part by the Korea
Research Foundation.

\section*{Appendix}
Making use of the identity
\begin{eqnarray}
\int\frac{t[-(t+l)^2+\beta]}{[(t+l)^2+\beta]^2}dt
&=&-\frac{l(t+l)+\beta}{(t+l)^2+\beta}-\frac{1}{2}\ln[(t+l)^2+\beta]
\nonumber\\
&&\mbox{ for}~~~ (t+l)^2+\beta>0,
\end{eqnarray}
one can find the following equality which could be used to find (11)
and (12) from (10);
\begin{eqnarray}
&&\sum_{l=1}^\infty\int_{-\frac{1}{2}}^\frac{1}{2}
	 \frac{t[-(t+l)^2+x-\frac{1}{4}]}
		  {[(t+l)^2+x-\frac{1}{4}]^2}dt \nonumber\\
&&=\frac{1}{4x}+\frac{1}{2}\ln x
+\lim_{N\rightarrow\infty}
   \{\sum_{n=1}^{N-1}\frac{n+\frac{1}{2}}{(n+\frac{1}{2})^2+x-\frac{1}{4}}
	 -\frac{1}{2}\ln[(N+\frac{1}{2})^2+x-\frac{1}{4}]\}.
\end{eqnarray}

\end{document}